\documentclass[12pt]{article}
\usepackage{epsf,amsfonts,amssymb,epsfig,amsmath,graphics}
\usepackage{hyperref}
\renewcommand{\text}[1]{#1}

\newcommand{\be}{\begin{equation}}
\newcommand{\ee}{\end{equation}}
\newcommand{\ben}{\begin{displaymath}}
\newcommand{\een}{\end{displaymath}}
\newcommand{\bea}{\begin{eqnarray}}
\newcommand{\eea}{\end{eqnarray}}
\newcommand{\bean}{\begin{eqnarray*}}
\newcommand{\eean}{\end{eqnarray*}}
\newcommand{\nn}{\nonumber \\}
\newcommand{\ba}{\begin{array}}
\newcommand{\ea}{\end{array}}
\newcommand{\bi}{\begin{itemize}}
\newcommand{\ei}{\end{itemize}}

\def\s{\sigma}

\begin{document}

\makeatletter
\renewcommand{\theequation}{\thesection.\arabic{equation}}
\@addtoreset{equation}{section}
\makeatother

\baselineskip 18pt

\begin{titlepage}

\vfill

\begin{flushright}
Imperial/TP/2010/JG/02\\
\end{flushright}

\vfill

\begin{center}
   \baselineskip=16pt
   {\Large\bf  Lifshitz Solutions of
 D=10 and D=11 supergravity}
  \vskip 1.5cm
      Aristomenis Donos$^1$ and Jerome P. Gauntlett$^{1,2}$\\
   \vskip .6cm
      \begin{small}
      \textit{$^1$Blackett Laboratory, 
        Imperial College\\ London, SW7 2AZ, U.K.}
        \end{small}\\*[.6cm]
   \vskip .3cm
      \begin{small}
      \textit{$^2$Perimeter Institute for Theoretical Physics\\
        Waterloo, Ontario, N2L 2Y5, Canada}
        \end{small}\\*[.6cm]

\end{center}

\vfill

\begin{center}
\textbf{Abstract}
\end{center}

\begin{quote}
We construct infinite families of Lifshitz solutions of 
$D=10$ and $D=11$ supergravity with dynamical exponent $z=2$.
The new solutions are based on five- and seven-dimensional 
Einstein manifolds and are dual to field theories with Lifshitz scaling in 1+2 and 
1+1 spacetime dimensions, respectively. When the Einstein spaces
are Sasaki-Einstein, the solutions are supersymmetric.
\end{quote}

\vfill

\end{titlepage}
\setcounter{equation}{0}


\section{Introduction}
It is a tantalising possibility that holographic techniques in string/M-theory can
be used to  study strongly coupled condensed matter systems
using weakly coupled theories of gravity. One focus has been on 
phase transitions that exhibit an anisotropic scaling of time and space:
\be\label{lifsc}
t\to \lambda^z t,\qquad x^i\to \lambda x^i
\ee
where $z$ is called the dynamical exponent. The case $z=1$ 
arises when the critical point has full conformal invariance,
and one might hope to model it using a solution of 
string/M-theory with an anti-de-Sitter ($AdS$) factor.

There have been two main approaches in trying to extend holographic technology
to model field theories with $z\ne 1$.
In the first approach, one generalises the 
$AdS$ geometry to a Lifshitz($z$) geometry of the form
\cite{Kachru:2008yh} (see also \cite{kl})
\be
ds^{2}_{Lif}=-r^{2z}\,dt^{2}+r^{2}\,d\vec{x}^{2}+\frac{dr^{2}}{r^{2}}
\ee
The scaling invariance \eqref{lifsc} is achieved by also scaling
the holographic coordinate $r\to \lambda^{-1}r$. In addition
these geometries are also invariant under time translations and 
spatial translations and rotations.
In the second approach, one instead considers a Schr\"odinger($z$) geometry
of the form \cite{Son:2008ye}\cite{Balasubramanian:2008dm}:
\be
ds_{Sch}^{2}=-r^{2z}\,(dx^+){}^{2}+2r^{2}\,dx^+dx^{-}+r^{2}\,d\vec{x}^{2}+\frac{dr^{2}}{r^{2}}
\ee
Here $x^+$ is identified with the dual field theory time coordinate $t$ and
the scaling \eqref{lifsc} is achieved by also transforming the
``extra" holographic coordinate $x^-$ as $x^-\to \lambda^{2-z}x^-$.
These geometries have additional symmetries to those of Lifshitz($z$).
There are Killing vectors which generate non-relativistic
boosts in the dual theory and the translation invariance of $x^-$
is associated with conservation of particle number.
In the special case that $z=2$ the
symmetry algebra is enlarged even further to include special conformal transformations
and the full symmetry algebra is the Schr\"odinger algebra. 
It should be noted that compared to the AdS case the holographic dictionary 
for both Lifshitz and Schr\"odinger geometries is still very much in its infancy.

The initial constructions of these geometries have been in the context of ``bottom up" 
models. In this approach one advocates a simple theory of gravity, typically in $D=4$
or $D=5$ dimensions, coupled to a small number of matter fields
with some simple couplings. The main virtue of this
approach is that one can easily start 
to investigate possible scenarios where
string/M-theory might lead to new insights into condensed matter.
One hopes that the solutions one finds exist somewhere in the landscape 
of string/M-theory (or perhaps provide a good approximation to such solutions)
and hence correspond to bone-fide dual field theories.  However, it is not clear if 
these hopes are realised, and the bottom up results could be misleading.

The alternative ``top-down" approach aims at constructing holographic
solutions directly in string/M-theory.
In practise the focus is to construct
solutions of $D=10/11$ supergravity\footnote{In certain limits, one can also consider
probe branes in supergravity backgrounds.}, possibly leaving
issues such as perturbative and non-pertubative stability to future work.
The main advantage of the top down approach is that the relevant
solutions of string/M-theory will in fact correspond
to dual field theories. Another advantage is that, somewhat surprisingly, 
it is often the case  that one can explore
infinite classes of solutions at the same time. 
Moreover, within these infinite classes one can find universal phenomena.
The only disadvantage of the top down approach is that it is much 
harder to construct the solutions, 
and maybe harder still to construct solutions of
direct relevance to condensed matter systems. 

The original constructions of the Lifshitz and Schr\"odinger geometries were
in the context of simple bottom up models. In the Schr\"odinger case top down
solutions were subsequently constructed using
duality transformations and/or consistent Kaluza-Klein 
reductions \cite{Herzog:2008wg}\cite{Maldacena:2008wh}\cite{Adams:2008wt}.
These string/M-theory solutions have been significantly further generalised 
in \cite{Gauntlett:2009zw}-\cite{Jeong:2009aa}.

By contrast, it has proved surprisingly 
difficult to construct top-down Lifshitz solutions. 
Indeed the difficulties led to the work of 
\cite{Li:2009pf} which proved a no-go theorem concerning their existence 
(see also \cite{Blaback:2010pp}).
However, 
in \cite{Hartnoll:2009ns} three schematic constructions of top down solutions
were discussed. More recently Balasubramanian and Narayan 
\cite{Balasubramanian:2010uk} have found 
Lifshitz solutions of type IIB and $D=11$ supergravity with $z=2$.
In this paper we will clarify and 
substantially generalise the solutions of \cite{Balasubramanian:2010uk}.

Remarkably, our new Lifshitz solutions, all of which have $z=2$, can be (essentially) obtained
from the same general class of type IIB and
D=11 supergravity solutions that were used to obtain Schr\"odinger($z$) solutions
(for various $z$) in \cite{dgsch2}. The solutions are constructed using 
five- and seven-dimensional Einstein spaces and for the special case when the Einstein space is Sasaki-Einstein, 
the solutions are supersymmetric, generically preserving two supersymmetries. As we shall see, 
our new solutions are closely related to 
the construction of Schr\"odinger($z)$ solutions of \cite{dgsch2} with $z=0$.

In this paper we will use the notation that a Lif$_D$($z$) solution is one
with $D$ non-compact directions that is dual to a field theory in $D-1$
spacetime dimensions. The solution must have the usual time and space translations,
spatial rotations and Lifshitz scaling with dynamical exponent $z$.
In this notation a Lif$_D$($z=1$) solution with conformal invariance
is the same as an $AdS_D$ solution. 
In general the solutions that we construct will
have the property that the $D$-dimensional non-compact part of the geometry
will depend on the coordinates of the internal compact dimensions\footnote{Note that this is
not possible in the case of $AdS_D$ solutions. Also note that this means that one can have a $Lif_D(z=1)$ solution of D=10/11 supergravityÊ
that is not an $AdS_D$ solution.}. Because
of this the solutions should be viewed directly in $D=10/11$. However, our solutions
include examples where the non-compact part of the geometry is independent
of the internal coordinates and the solutions can also be obtained
as solutions of a reduced $D$-dimensional theory of gravity. 
Developing the holographic
dictionary for the latter class of solutions should be easier than for the former class. 

In section 2, using arbitrary five-dimensional Einstein spaces, $E_5$, we construct 
Lif$_{4}$($z=2)$ solutions in type IIB supergravity. For a special sub-class we 
can T-dualise and uplift to obtain Lif$_{4}$($z=2)$ solutions of type IIA and D=11
supergravity, respectively. The most general class of solutions
exist when $E_5$ has non-vanishing second Betti number and
we illustrate with some explicit examples that include the Sasaki-Einstein spaces $T^{1,1}$ 
\cite{Candelas:1989js} and $Y^{p,q}$ \cite{Gauntlett:2004yd}. 

Our constructions include solutions in $D=11$ that
consist of a direct product of a Lif$_{4}$($z=2)$ 
factor  with a seven-dimensional compact manifold, with the latter
a two-torus fibred over $E_5$.
Moreover, the  Lif$_{4}$($z=2)$ factor is independent of the coordinates of the compact manifold.
The reason that these solutions evade the no-go theorem of
\cite{Li:2009pf} is simply that this theorem does not cover the most
general type of four-form flux.

In section 3 we present analogous constructions of 
Lif$_{3}$($z=2)$ solutions of D=11 supergravity
using seven-dimensional Einstein spaces, $E_7$. 
These new solutions again extend those discussed in
\cite{Balasubramanian:2010uk}.
We briefly conclude in section 4. 

The paper contains two appendices. In appendix A we carry out 
a dimensional reduction of type IIB on $S^1\times E_5$ and of $D=11$
on $S^1\times E_7$
to obtain constrained theories of gravity in
$D=4$ and $D=3$, respectively, from which we can make contact with the bottom
up constructions of the Lifshitz($z$) solutions. For the special case when
$E_5=T^{1,1}$ we show that there is a more elaborate and consistent Kaluza-Klein 
reduction on $S^1\times T^{1,1}$. In appendix B we 
we provide an alternative verification of the supersymmetry of the Lif$_4$($z=2$) solutions of $D=11$ supergravity.

\section{Lif$_{4}$($z=2$) Solutions}\label{seciib}
Our starting point is the following 
ansatz (essentially as in \cite{dgsch2}) 
for the bosonic fields of type IIB supergravity:
\bea\label{start1}
ds_{10}^{2}&= & \Phi^{-1/2}\left[2dx^+dx^-+h(dx^+)^2 
+dx_1^2+dx_2^2     \right] +\Phi^{1/2}ds^2(M_6)\nn
F_{5}&= & dx^{+}\wedge dx^{-}\wedge dx_{1}\wedge dx_{2}\wedge d\Phi^{-1}
+\,\ast_{M}d{\Phi}\nn
G&= & dx^{+}\wedge W\nn
P&=&gdx^+
\eea
where $G$ is the complex three-form and the complex one-form $P$ 
incorporates the axion and dilaton\footnote{Our 
conventions for type IIB supergravity \cite{Schwarz:1983qr,Howe:1983sra}
are as in \cite{Gauntlett:2005ww}. In particular, 
packaging the dilaton, $\phi$, and the axion, $C_0$,
as $\tau=C_0+ie^{-\phi}$, then 
$P=(i/2)e^\phi d\tau$ and $G=ie^{\phi/2}(\tau dB-dC_2)$
where $B$, $C_2$ are the NS and RR 2-form potentials, 
respectively.}.
Here $\Phi$, $h$, $g$ are functions and $W$ is a 
complex two-form all defined on the six-dimensional Ricci-flat
manifold, $M_6$, and they can all have a functional dependence
on the coordinate $x^+$. 
One finds that all the equations of motion are satisfied provided that
\begin{align}\label{begeqs1}
\nabla^{2}_{M}\Phi&=0\nn
dx^+\wedge dW=d\ast_{M}W & =0\nn
\nabla^{2}_{M}h & =-4g^2\Phi-|W|^2_{M}\end{align}
where $|W|^2_{M}\equiv (1/2!)W^{ij}W^\ast_{ij}$ with indices raised
with respect to the metric on $M_6$.
Observe that when $h=W=0$ we have the standard D3-brane class
of solutions with a Ricci-flat transverse space.
As we will review a little later, when $M_6$ is a Calabi-Yau 3-fold,
$M_6=CY_3$, supersymmetry is preserved for certain choices of $W$ \cite{dgsch2}.
The general structure of these solutions is that of D3-branes transverse to
a Ricci-flat space with a wave propagating on the world-volume, and carrying 
additional RR and NS magnetic 3-form flux.

We now specialise to the case that $M_6$ 
is a metric cone over a five-dimensional compact Einstein manifold $E_5$,
$ds^2(M_6)=dr^2+r^2ds^2(E_5)$. The Einstein metric is normalised so that
its Ricci tensor is equal to four times the metric, the same as for a 
round five-sphere. When $M_6=CY_3$ then the Einstein manifold is Sasaki-Einstein.
In order to get solutions with Lif$_4$$(z=2)$ symmetry we now set
\bea
\Phi&=&r^{-4}\nn
h&=& r^{-2}f
\eea
where $f$ is a function of the coordinates on $E_5$ and 
$x^+$. In addition 
$W$ and $g$ are taken to be a three-form and a function defined on $E_5$ 
and they are both also functions of $x^+$. 
The equations \eqref{begeqs1} are now solved provided that
\begin{align}
dx^+\wedge dW =d\ast_{E}W  &=0\nn
-\nabla^2_{E}f+4f&=  4\left|g\right|^{2}+\left|W\right|^{2}_E
\label{eq:f_equation}
\end{align}
Observe that when $g=W=0$, necessarily we have $f=0$ since
eigenvalues of the Laplacian on compact $E_5$ are negative. For a similar 
reason, given $g$, $W$ any solution to the second equation is necessarily 
unique.

After relabelling 
\be
x^-=t,\qquad x^+=\sigma 
\ee
the full solution now reads
\begin{align}\label{eq:generalIIB_ansatz}
ds^{2}= & r^{2}\,\left[2d\s dt+dx_{1}^{2}+dx_{2}^{2}\right]+\frac{dr^{2}}{r^{2}}+f\,d\s^{2}+ds^{2}\left(E_5\right)\nn
=&-\frac{r^{4}}{f}\,dt{}^{2}
+r^{2}\,\left(dx_{1}^{2}+dx_{2}^{2} \right)+\frac{dr^{2}}{r^{2}}
+f\,\left(d\s+\frac{r^2}{f}dt \right)^{2}
+ds^{2}\left(E_5\right)\nn
F_{5}= & 4r^{3}\, d\s\wedge dt\wedge dr\wedge dx_{1}\wedge dx_{2}
+4\,\mathrm{Vol}_{E_5}\nn
G= & d\s\wedge W\nn
P= & g d\s\end{align}
with $f,g,W$ satisfying \eqref{eq:f_equation}.
Observe that when $f=g=W=0$ we have the standard $AdS_5\times E_5$ solution.
The solutions of \cite{Balasubramanian:2010uk} can be recovered in the
special case when 
$W=0$, $f$ and $g$ are functions only of $\s$ (i.e. are independent of
the coordinates of $E_5$) and furthermore that
$g$ is real (i.e. the axion is zero). 
By restricting to solutions with $f>0$, 
following \cite{Balasubramanian:2010uk}, we can view these
as Lif$_{4}$($z=2)$ solutions by taking $\s$ to parametrise a 
compact $S^1$. In particular, the full solution is invariant
under the following scalings of the four non-compact directions, 
parameterised by $t,x^1,x^2,r$:
\be 
t\to \lambda^2 t,\qquad x^i\to \lambda x^i,\qquad r\to \lambda^{-1}r
\ee
These solutions correspond to dual field theories in $d=3$ spacetime
dimensions with dynamical exponent $z=2$.

When $f$ has dependence on $\s$ and the coordinates 
on the Einstein space, a $D=4$ perspective of the solutions
is artificial and they should be viewed directly in $D=10$.
However, there is an interesting sub-class of solutions where $f$ is a constant,
and without loss of generality we can set $f=1$. These solutions
require $4=4|g|^2+|W|^2_E$, with $W$ a harmonic form on $E_5$,
and we will present some explicit examples with $W\ne 0$ 
below. For this class the $D=4$
non-compact part of the metric is precisely that of the original 
Lif$_{4}$($z=2$) geometry of \cite{Kachru:2008yh}. 
In appendix A we will make contact with the
bottom up construction of \cite{Kachru:2008yh} by performing a dimensional reduction
of type IIB supergravity on $S^1\times E_5$. 

For this class of solutions, and more generally for solutions where
$f$, $g$ and $W$ are independent of the $\s$ coordinate, which implies that
$W$ is a harmonic two-form on $E_5$, the 
vector $\partial_\s$ is a Killing vector that
also preserves the fluxes. This will generate a global symmetry
in the dual $d=3$ dimensional field theory (as will any
isometries of the Einstein space $E_5$ that also preserve $g$ and $W$).
The most general solutions will also have a dependence on $\s$ and
$\partial_\s$ will no longer be Killing.

It is worth highlighting that for the general class of solutions
given in \eqref{eq:generalIIB_ansatz},\eqref{eq:f_equation}, the vectors
$-x^+\partial_i+x^i\partial_-$, $i=1,2$ are also Killing, where for this
paragraph we have temporarily reverted $t,\s$ back to $x^-,x^+$, respectively.
When $x^+$ is non-compact these Killing vectors generate 
the finite transformation
\bea\label{boost}
x^i\to x^i-u^ix^+,\qquad
x^-\to x^-+ u^i x^i-\tfrac{1}{2}u^2x^+
\eea
for constant $u^i$. This symmetry is explicitly broken by taking $x^+$
to be compact\footnote{To see this, note that if $x^+\equiv x^++2\pi R$ then
$(x^+,x^-,x^i)$ and $(x^++2\pi R,x^-,x^i)$, which parametrise the same point, do not get mapped to the same point by the
finite transformation.}.
In fact when $x^+$ is
non-compact the solutions can be viewed as
$z=0$ Schr\"odinger solutions of the type studied in \cite{dgsch2}
with $x^+$ playing the role of the
time coordinate and $x^-$ the auxiliary coordinate 
(leading to conservation of
particle number in the dual field theory). The transformation
\eqref{boost} then corresponds to the usual non-relativistic boosts.
By contrast, here we have obtained
Lif$_{4}$($z=2$) solutions by switching the roles of $x^+$ and $x^-$ and then
compactifying $x^+$. 

It is curious that for the new
Lif$_{4}$($z=2$) solutions given in \eqref{eq:generalIIB_ansatz} 
the Killing vector $\partial_t$ is always null;
this won't be the case for the type IIA solutions which we obtain after 
T-duality and are presented after the following discussion of supersymmetry.

\subsection{Supersymmetry}
When the Einstein space is taken to be Sasaki-Einstein, 
or equivalently the Ricci-flat cone $M_6$ is $CY_3$,
the Lif$_{4}$($z=2$) solutions that we have presented can preserve
supersymmetry. In fact this follows from the analysis of \cite{dgsch2}.
Specifically, if the two-form $W$ is of type
$(1,1)$ and primitive or of type $(0,2)$ on the $CY_3$ cone then
the more general solutions \eqref{start1}
generically preserve 2 supersymmetries.
More specifically, for the $D=10$ metric in \eqref{start1}
we introduce the frame 
$e^+=\Phi^{-1/4}dx^+, e^-=\Phi^{-1/4}(dx^-+\frac{h}{2}dx^+), 
e^2=\Phi^{-1/4}dx^1,$ etc. and
choose positive orientation to be given by $e^{+-23}\wedge\rm{Vol}_{CY}$, where $\rm{Vol}_{CY}$ is the volume element on $CY_{3}$.
Consider first the special case where $g=h=W=0$. Then, as usual,
a generic $CY_3$ breaks 1/4 of the supersymmetry, 
while the harmonic function $\Phi$ leads
to a further breaking of 1/2, the Killing spinors satisfying the additional projection $\Gamma^{+-23}\epsilon=i\epsilon$ (with the Killing
spinors gaining a factor $\Phi^{-1/8}$).
Switching on $g,h,W$ we find that we need to also impose $\Gamma^+\epsilon=0$ 
and $\Gamma^{ij}W_{ij}\epsilon^c=0$ (and the spinors gain a dependence on $x^+$).
Note that in the supersymmetric Lif$_{4}$($z=2$)
solutions, $W$ is a two-form on the $SE_5$ space and hence $W$ should be $(1,1)$ on the cone,
or, equivalently, $(1,1)$ with respect to the local four-dimensional K\"ahler-Einstein
base space associated with the $SE_5$.

We can also determine the supersymmetry algebra by constructing the 
Killing vector that arises from bi-linears of these Killing
spinors. Specifically, following \cite{dgsch3} we find that if we define
\be
K^M=\bar\epsilon\Gamma^M\epsilon
\ee
then $K=\partial_t$. In other words the supersymmetry is squaring to 
the Killing vector generating time translations in the dual Lifshitz field
theory. As noted above, for these type IIB solutions
this Killing vector is null.

\subsection{Type IIA and $D=11$ pictures}
Starting with the type IIB solutions given in \eqref{eq:generalIIB_ansatz}
we can obtain analogous solutions of type IIA by performing 
a T-duality on the $\s$ direction. To do this we require that
$f$ and $W$ are independent of $\s$. We also require $g$ to be independent
of $\s$ which means that the dilaton and axion are constant and 
for simplicity we take them to be trivial so that $G=-(dB^{(2)}+idC^{(2)})$. 
The function $f$ satisfies
\be\label{theeqagain}
4f-\nabla^2_Ef=|W|^2_E
\ee

Writing
\be
W=d{A}^{\left(1\right)}+i\,d{A}^{\left(2\right)},\qquad D\s\equiv d\s-{A}^{\left(1\right)}
\ee
and performing the T-duality we obtain the type IIA solutions
\begin{align}\label{iiasol}
ds^{2}=&-\frac{r^{4}}{f}dt{}^{2}+r^{2}\,\left(dx_{1}^{2}+dx_{2}^{2}\right)+\frac{dr^{2}}{r^{2}}+\frac{1}{f}\,D\s^{2}+ds^{2}\left(E_5\right)\nn
e^{2\phi}=&\frac{1}{f}\nn
B=&-D\s\wedge \frac{r^2}{f}\,dt\nn
F_{2}=&-d{A}^{\left(2\right)}\nn
F_{4}=&4r^{3}\,dt\wedge dx_{1}\wedge dx_{2}\wedge dr+D\s\wedge \frac{r^{2}}{f}dt\wedge d{A}^{\left(2\right)}
\end{align}
where $F_4$ is the type IIA RR four-form field strength satisfying $dF_4=H_{3}
\wedge F_{2}$.
It is interesting to observe that the Killing vector $\partial_t$, generating
the time-translations of the dual Lifshitz field theory, is now time-like.
This is to be contrasted with the IIB solutions where it was null.
We also observe that the $\s$ circle direction is now, in general, non-trivially fibred over 
the Einstein space. In particular, 
the first Chern class of this circle bundle is given by
the cohomology class of $d{A}^{\left(1\right)}$ and this will lead, in general, 
to the six-dimensional internal space no longer having the topology of
$S^1\times E_5$.

We can uplift these solutions to $D=11$ on a circle parametrised by $\chi$
and we obtain
\begin{align}
ds^{2}=&f^{1/3}\left[-\frac{r^{4}}{f}\,dt^{2}
+r^{2}\left(dx_{1}^2+dx_{2}^{2}\right)
+\frac{dr^{2}}{r^{2}}\right]
+f^{-2/3}\left[D\s^2+D\chi^{2}\right]
+f^{1/3}ds^2\left(E_{5}\right) \nn
G_{4}=&dt\wedge d\left(r^{4}dx_{1}\wedge dx_{2}+\frac{r^{2}}{f}\, D\chi\wedge D\s\right)
\label{eq:11D_susy}
\end{align}
where $D\chi\equiv d\chi-A^{(2)}$.
It is illuminating to rewrite these $D=11$ solutions in the form arising from performing
a T-duality and then uplifting the more general solutions \eqref{start1}:
\bea\label{d11solf}
ds^2&=&-H_1^{-2/3}H_2^{-2/3}dt^2
+H_1^{-2/3}H_2^{1/3}\left(dx_{1}^2+dx_{2}^{2}\right)
+H_1^{1/3}H_2^{-2/3}\left[D\s^2+D\chi^{2}\right]\nn
&&\, \qquad \qquad\qquad\qquad+H_1^{1/3}H_2^{1/3}\left[dr^2+r^2ds^2(E_5)\right]\nn
G_4&=&
dt\wedge dx_{1}\wedge dx_{2}\wedge d(H_{1}^{-1})
+dt\wedge d( H_{2}^{-1}\,D\chi\wedge D\s)
\eea
In the Lif$_4$($z=2$) solutions $H_1=r^{-4}$, $H_2=fr^{-2}$. The general structure of the solution is that
of two membranes intersecting in the time direction: the world-volume of one of
the membranes is where the dual field theory resides and the other membrane
is wrapped over a two-torus, parametrised by $\s,\chi$, which is fibred over 
an overall transverse six-dimensional Ricci-flat space. Note that the NS and RR 3-form flux
in the type IIB picture now manifests itself through the fibration.

For the special case that the Einstein space is Sasaki-Einstein, 
these type IIA and $D=11$ solutions will preserve supersymmetry provided 
that $W$
is of type (1,1) and primitive on the $CY_3$ cone. 
This follows because the type IIB Killing spinors are independent of the $\s$ direction
and hence are preserved under the T-duality transformation. The 
Killing vector that can be constructed from the generic $D=11$ Killing spinors 
is again $\partial_t$ which, as we have noted above, is now timelike.
We provide an independent derivation of this in appendix \ref{susystructure}.

For the special case when $f=1$ (which can occur, for example,
when $E_5=T^{1,1}$ as we show below), this solution
is a direct product of a Lif$_4$($z=2)$ geometry with a compact 
seven-dimensional internal space, the latter a two-torus fibration over $E_5$. The reason that the
solutions evade the no-go theorem of \cite{Li:2009pf} is simply because
\cite{Li:2009pf} did not consider the most general flux compatible with
Lif$_4$($z=2)$ symmetry.

\subsection{Explicit Examples}
Type IIB solutions with $W=0$ and $g\ne 0$ can be found for
any choice of $E_5$. Indeed for real $g$ (i.e. the axion is zero)
these solutions were constructed in \cite{Balasubramanian:2010uk}. 
More interesting solutions have $W\ne 0$. Since $W$ is harmonic
on $E_5$, the latter must have non-trivial two- and three-cycles. Thus there
are no such solutions on $S^5$. A simple non-supersymmetric solution
can be constructed for $E_5=S^2\times S^3$. Let us discuss in a bit more 
detail some supersymmetric examples using Sasaki-Einstein
spaces, first considering $E_5=T^{1,1}$ and then $E_5=Y^{p,q}$.

The metric for $T^{1,1}$ can be written
\be
ds^2(E_5)=\frac{1}{9}(d\psi-\cos\theta_1 d\phi_1-\cos\theta_2 d\phi_2)^2
+\frac{1}{6}(d\theta_1^2+\sin\theta_1 d\phi_1^2)
+\frac{1}{6}(d\theta_2^2+\sin\theta_2 d\phi_2^2)
\ee
For $W$ we choose 
\be
W=
\frac{k}{\sqrt{18}}(\sin\theta_1 d\theta_1\wedge d\phi_1-\sin\theta_2 d\theta_2\wedge d\phi_2)
\ee
where $k=k(\s)$ and it is simple to check that it is harmonic (and thus 
closed and co-closed) on $T^{1,1}$.
The unique solution to \eqref{eq:f_equation} is given by
\be
f=k^2+|g|^2
\ee
and is independent of the coordinates on $T^{1,1}$
In particular, 
solutions that are independent of $\s$, which covers the type IIA
and $D=11$ solutions, have constant $f$.
It is straightforward to check that $W$ is of type (1,1) and primitive
on the cone over $T^{1,1}$ (the conifold), and hence this whole
family of solutions preserve supersymmetry. 
Note that this construction can be immediately
adapted to the Einstein spaces $T^{p,q}$ \cite{Romans:1984an}
to obtain analogous non-supersymmetric solutions.

We next turn to examples where $E_5=Y^{p,q}$. The Sasaki-Einstein metric
can be written in the canonical form \cite{Gauntlett:2004yd}
\be
ds^{2}(Y^{p,q})
=\frac{1}{9}\left(d\psi^{\prime}+\sigma\right)^2+ds^{2}_{4}
\ee
where 
\be
ds^{2}_{4}=\frac{1-y}{6}\,\left(d\theta^{2}+\sin^{2}\theta\,d\phi^{2}\right)+\frac{dy^{2}}{w\left(y\right)q\left(y\right)}+\frac{1}{36}w\left(y\right)q\left(y\right) \left(d\beta+\cos\theta d\phi\right)^2\nonumber\\
\ee
is a {\it locally} defined K\"ahler-Einstein metric, and
\bea
\sigma=&y\left(d\beta+\cos\theta\,d\phi \right)-\cos\theta d\phi
\eea
and
\begin{align}
w\left(y\right)=\frac{2\left(a-y^2\right)}{1-y},\qquad
q\left(y\right)=\frac{a-3y^2+2y^3}{a-y^2}
\end{align}
For $W$ we choose 
\begin{equation}
W=\frac{1}{\sqrt{72}}\,d\left[\frac{1}{1-y}\,\left(d\beta+\cos\theta\, d\phi \right)\right]
\end{equation}
It can easily be checked that $W$ is co-closed on $Y^{p,q}$ and
furthermore, that
it is $\left(1,1\right)$ and primitive with respect to the local 
four-dimensional KE metric $ds^2_{4}$ and hence is also on
the corresponding $CY_3$. To see that this two-form 
is globally defined on the whole Sasaki-Einstein it is helpful to use 
the coordinates defined by
\be
\alpha=-\frac{1}{6}(\beta+\psi'),\qquad \psi=\psi'
\ee
In these coordinates the metric can be written
\be
ds^2=w(y)(D\alpha)^2+\frac{dy^2}{w(y)q(y)}+\frac{q(y)}{9}D(\psi)^2+
\frac{1-y}{6}(d\theta^2+\sin^2\theta d\phi^2)
\ee
where 
\begin{align}
D\alpha=d\alpha+\frac{a-2y+y^2}{6\left(a-y^2\right)} D\psi,\qquad
D\psi=d\psi-\cos\theta d\phi
\end{align}
As discussed in detail in \cite{Gauntlett:2004yd}, in these coordinates 
one can show that there is a circle fibration, parametrised by $\alpha$ 
over a globally defined four-dimensional base. This base space, in turn,
is a two-sphere fibration, parametrised by $\psi,y$ with $y_1\le y\le y_2$
where $y_i$ are two suitable roots of $q(y)$, 
over the round two-sphere, parametrised
by $(\theta,\phi)$. In this construction $a$ and the roots $y_i$ are fixed by
two relatively prime integers $p>q>0$:
\bea
a&=&\frac{1}{2}-\frac{p^2-3q^2}{4p^3}\sqrt{4p^2-3q^2}\nn
y_1&=&\frac{1}{4p}\left(2p-3q-\sqrt{4p^2-3q^2}\right)\nn
y_2&=&\frac{1}{4p}\left(2p+3q-\sqrt{4p^2-3q^2}\right)
\eea
In this construction the one-form $D\alpha$ is the globally defined
one-form of the circle fibration. Furthermore
the two-form $dy\wedge D\psi$ is also globally defined. 
Writing $W$ in the new 
coordinates we conclude that it is globally defined.

\begin{figure}[htb!]
\centering%
\includegraphics[scale=0.75]{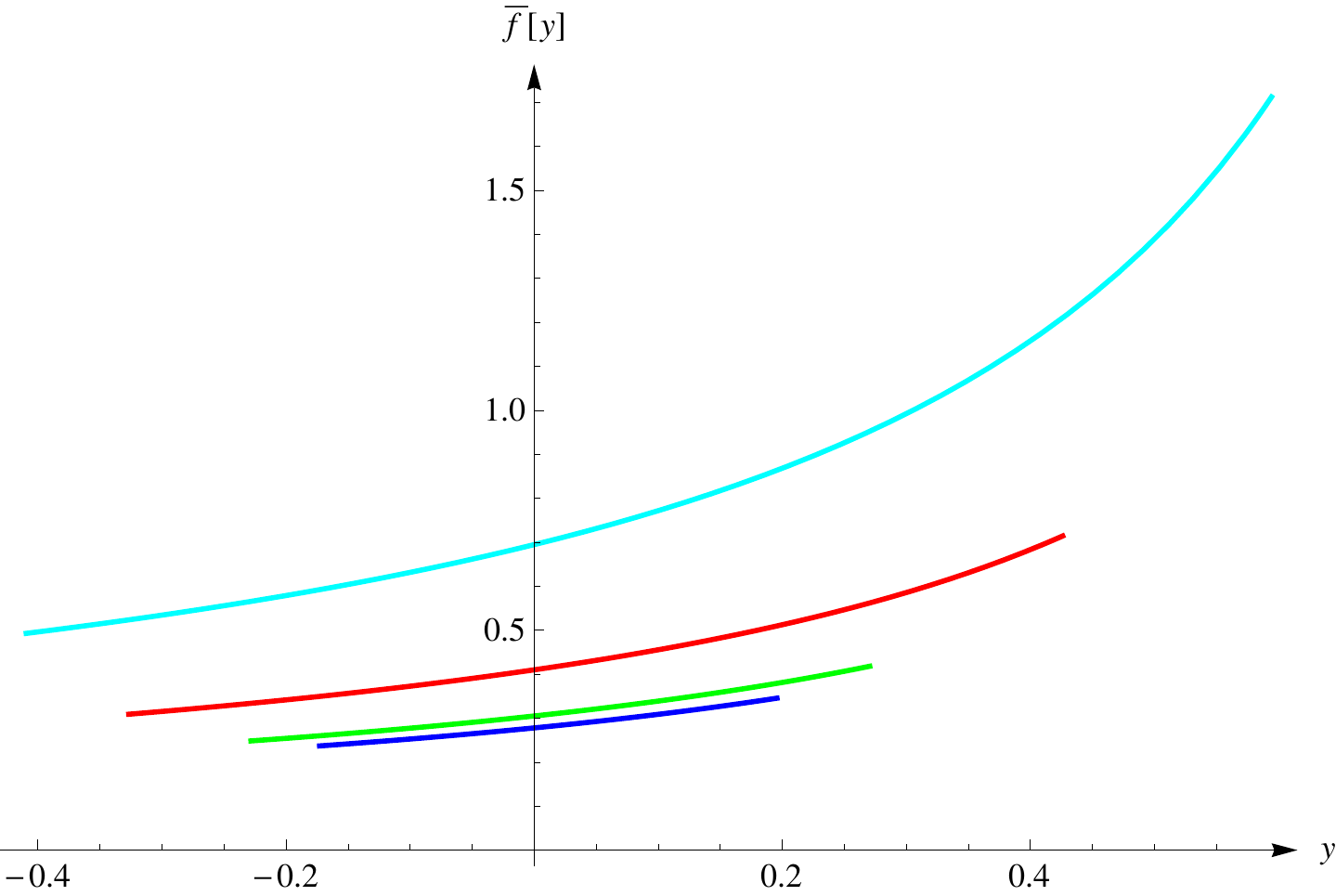}
\caption{Numerical solutions for the function $\bar f(y)$ 
for $(p,q)$= $(4,1)$ (i.e. $a\sim 0.10$) - dark blue;
$(3,1)$ (i.e. $a\sim 0.18$) - green;
$(2,1)$ (i.e. $a\sim 0.39$) - red; $(3,2)$ (i.e. $a\sim 0.64$) - cyan.
For each case we have plotted $\bar f$ just for the
values $y\in[y_1,y_2]$.}
\label{fig:Numerical_Ypq}
\end{figure}

The general solution to \eqref{eq:f_equation} will be given by 
$f=\bar f+\left|g\right|^{2}$ where $\bar f$ satisfies 
\begin{equation}
-4\bar f+\frac{2}{1-y}\,\left[\left(a-3y^2+2y^3 \right)\bar f^{\prime} 
\right]^{\prime}+\frac{1}{\left(1-y\right)^4}=0
\end{equation}
If $g=0$ (as in the type IIA and $D=11$ solutions) 
it is important that $\bar f$ (and hence $f$) is
strictly positive in the interval $y_1\le y\le y_2$. 
We have not found an analytic expression for the function $\bar f$ 
but we numerically solved it for a few values of $a$, for specific
values of $p$ and $q$, as shown in 
figure \ref{fig:Numerical_Ypq}.
Note that $\bar f$ monotonically increases with $a$ and also that
it diverges as $a\to 1$, which is expected since $a=1$ is the case
of $S^5$, where there are no solutions with $W\ne 0$.

It would be interesting to generalise these solutions by replacing $Y^{p,q}$ with the
more general $L^{a,b,c}$ Sasaki-Einstein metrics of \cite{Cvetic:2005ft}.
In the above examples, where the topology of $E_5$ is $S^2\times S^3$, we have
$H^2(E_5,\mathbb{Z})=\mathbb{Z}$ and
hence the circle bundle over $E_5$ appearing in the type IIA metric in 
\eqref{iiasol} is specified by an integer $n$, the Chern number. Taking
$n=\pm 1$ gives a total space of topology $S^3\times S^3$ (taking it to be $n\ne \pm 1$ would instead lead to
a non-simply connected total space, which can always be lifted to the simply connected cover with
$n=\pm 1$). This lifts to $D=11$ solutions \eqref{d11solf}
with internal space of topology $S^3\times S^3\times S^1$.

\section{Lif$_{3}$($z=2$) Solutions}\label{d=11}

We consider the ansatz (essentially as in
\cite{dgsch2}) for the bosonic fields of $D=11$ supergravity given by
\bea\label{dipsy}
ds^{2}&= & \Phi^{-{2}/{3}}\left[2dx^{+}dx^{-}+h\,\left(dx^{+}\right)^{2}
+dx^{2}\right]+\Phi^{{1}/{3}}ds^{2}\left(M_8\right)\nn
G&= & dx^{+}\wedge dx^{-}\wedge dx\wedge d{\Phi}^{-1}+dx^{+}\wedge V
\eea
where $\Phi$, $h$ are functions and $V$ is a three-form 
on the eight-dimensional Ricci-flat space $M_8$, and they can all
have a dependence on the coordinate $x^+$.
Our conventions for $D=11$ 
supergravity \cite{Cremmer:1978km} are as in \cite{gp}.
One finds that all the equations of motion are satisfied provided that
\bea
\label{begeqs}
\nabla^{2}_{M}\Phi&= & 0\nn
dx^+\wedge dV=d\ast_{M}V&= & 0\nn
\nabla^{2}_{M}h&= & -|V|^2_{M}
\eea
where $|V|^2_{M}\equiv (1/3!)V^{ijk}V_{ijk}$ with indices raised
with respect to the metric on $M_8$.
When $h=V=0$ we have the standard M2-brane class
of solutions with a transverse Ricci-flat space $M_8$. 
When $M_8$ is a Calabi-Yau four-fold, $M_8=CY_4$, 
supersymmetry is preserved
for certain choices of $V$ \cite{dgsch2} , as we discuss below.
The general structure of these solutions is that of M2-branes transverse
to the Ricci-flat space with a wave propagating on the world-volume, that in addition carry
magnetic four-form flux.

We now specialise to the case that $M_8$ is a metric cone over a 
seven-dimensional Einstein manifold $E_7$,
$ds^2(M_8)=dr^2+r^2ds^2(E_7)$. The Einstein metric is normalised
so that its Ricci tensor is equal to six times the metric, the
same as for a round seven-sphere.
When $M_8=CY_4$ the Einstein manifold is Sasaki-Einstein $SE_7$.
In order to get solutions with 
Lif$_3$($z=2$) symmetry we now set
\bea
\Phi&=&r^{-6}\nn
h&=& fr^{-4}
\eea
where $f$ is a function of the coordinates on the Einstein space and 
$x^+$. In addition $V$ is taken to be a three-form defined on $E_7$ and a 
function of $x^+$. The equations \eqref{begeqs} are solved provided that
\bea\label{eeqns}
dV=d\ast_{E}V&= & 0\nn
8f-\nabla^2_Ef&=&|V|^2_E
\eea
For a given $V$, any solution of the second equation
is unique. In particular, if $V=0$ then $f=0$.
We note that this ansatz is similar to that considered in
appendix A.1 of \cite{Balasubramanian:2010uk}, however their ansatz 
implicitly assumes that $|V|^2=constant$ and $f$ did not have any dependence on
the Einstein space.

Introducing the new coordinates
\be
\rho=r^2,\qquad x\to x/2,\qquad x^+= \s/2,\qquad x^-= t/2
\ee
the full solution reads
\bea\label{fullsol}
ds^{2}&= & 
\frac{1}{4}\left[\rho^2\left(2d\s dt+dx^2\right)+\frac{d\rho^2}{\rho^2}+fd\s^2
\right]
+ds^{2}\left(E_{7} \right)\nn
&=&
\frac{1}{4}\left[-\frac{\rho^4}{f}dt^2+\rho^2 
dx^2+\frac{d\rho^2}{\rho^2}\right]
+\frac{f}{4}\left[d\s+\frac{\rho^2}{f}dt\right]^2 
+ds^{2}\left(E_{7} \right)\nn
G&=& \frac{3}{8}\rho^2\,d\s\wedge dt\wedge dx\wedge d\rho
+\frac{1}{2}d\s\wedge V
\eea
with $f,V$ satisfying \eqref{eeqns}.
Observe that when $f=V=0$ we have the standard $AdS_4\times E_7$ solution.

To view these as Lif$_3$($z=2)$ solutions we
consider solutions with $f>0$ and take the $\s$ coordinate 
to parametrise a compact internal $S^1$. The full solution is then
invariant under Lifshitz scalings of the three non-compact directions,
parametrised by $(t,x,r)$, with dynamical exponent $z=2$. In appendix
\ref{a3} we show how this
special sub-class of solutions can be obtained from a simple
theory of gravity in $D=3$ after dimensional reduction on $S^1\times E_7$.

When $f$ has dependence on $\s$ and/or the $E_7$ space, the solutions
should be viewed directly in $D=11$. An interesting sub-class has
constant $f$ and, without
loss of generality, we can take $f=1$ and $|V|^2=8$. 
These solutions, and more generally solutions that are independent of
$\s$, will have $\partial_\s$ as a Killing vector. These solutions
also have $-\s\partial_x+x\partial_t$ as a Killing vector and the 
corresponding finite transformation (see \eqref{boost}) is broken by
$\s$ being compact. As in the type IIB case discussed in section 3, in a 
certain sense, these solutions are closely related to the Schr\"odinger($z$)
solutions of \cite{dgsch2} with $z=0$.
We also observe that the Killing vector
$\partial_t$ is null in the full $D=11$ solution. However, for
the solutions independent of $\s$ we can dimensionally reduce on
$\s$ to obtain a type IIA solution where $\partial_t$ is time-like.

Some simple explicit examples 
can be obtained by taking $E_7=S^3\times E_4$
where $S^3$ is the round three-sphere and
$E_4$ is an arbitrary four-dimensional Einstein space
and $V$ is taken to be the volume form on the $S^3$.
We will postpone further constructions of explicit solutions to
future work.

\subsection{Supersymmetry}
When $E_7$ is taken to be Sasaki-Einstein, or equivalently $M_8$ 
is taken to be a Calabi-Yau four-fold, these Lif$_3$($z=2$) solutions can 
preserve supersymmetry \cite{dgsch2}. In particular,
if we choose the three-form $V$ to only have $(2,1)$ plus $(1,2)$ pieces 
and be primitive on the $CY_4$
then the more general solutions given in \eqref{dipsy} generically preserve 2 supersymmetries \cite{dgsch2}.
More specifically, we introduce the frame
$e^+=\Phi^{-1/6}dx^+, e^-=\Phi^{-1/6}(dx^-+\frac{h}{2}dx^+), e^2=\Phi^{-1/6}dx,$ etc. and
choose positive orientation to be given by $e^{+-2}\wedge\rm{Vol}_{CY}$, where $\rm{Vol}_{CY}$ is the volume element on $CY_{4}$.
Consider first the special case that $h=V=0$. Then, as usual,
a non-flat $CY_4$ breaks 1/8 of the supersymmetry, and the harmonic function $\Phi$ can be added ``for free''
(the projection on the Killing spinors arising from the $CY_4$ automatically imply the projection $\Gamma^{+-2}\epsilon=-\epsilon$).
Switching on $h,V$ we find that we need to also impose $\Gamma^+\epsilon=0$ and $\Gamma^{ijk}V_{ijk}\epsilon=0$.
As usual the skew-whiffed solutions, obtained by changing the sign of the four-form flux,  generically don't preserve any supersymmetry 
(apart from the special case when $SE_7=S^7$).

The supersymmetry algebra can be obtained by constructing the
Killing vector arising as bi-linears in the Killing spinors.
Following \cite{dgsch2} we find that the anticommutator
of the supersymmetries gives the null Killing vector $\partial_t$,

\section{Final Comments}
We have constructed rich classes of Lif($z$) solutions 
of $D=10$ and $D=11$ supergravity with $z=2$. This work opens up several
avenues for further exploration and we conclude by 
briefly mentioning some of them.

It will be interesting to see if our solutions can be simply 
generalised to Lifshitz solutions with other values of $z$. Another
direction will be to construct black hole solutions that asymptotically 
approach the new Lifshitz solutions. It will also be interesting to see
of it is possible to construct solutions that interpolate between 
Schr\"odinger, Lifshitz and $AdS$ geometries corresponding
to RG flows in the dual field theories. It has been proposed 
\cite{Gubser:2009cg} that Lifshitz geometries can naturally arise as the ground states
of holographic superconductors and it will be interesting to
see if our solutions appear in this way too.

\subsection*{Acknowledgements}
AD is supported by an EPSRC Postdoctoral Fellowship and
JPG is supported by an EPSRC Senior Fellowship and 
a Royal Society Wolfson Award.
We thank Jaume Gomis, Shamit Kachru, Nakwoo Kim, Dario Martelli, Simon Ross and Oscar Varela 
for helpful discussions. Research at Perimeter Institute is supported
by the Government of Canada through Industry Canada and by
the Province of Ontario through the Ministry of Research \& Innovation.

\appendix

\section{Reduced Theories of Gravity}

\subsection{Reduction of type IIB on $S^1\times E_5$ to $D=4$ when $f=1$}
Lifshitz solutions were first constructed in a bottom up
context in a $D=4$ theory of gravity with cosmological constant
coupled to a vector field and a two-form \cite{Kachru:2008yh}, 
or equivalently, to a single massive vector field \cite{Taylor:2008tg}. 
We show how the special class of type IIB solutions, for arbitrary
Einstein space, with $g=0$, $W$ independent of $\s$ and $f=1$, that we
constructed in section \ref{seciib}, are related to these constructions.
Specifically we dimensionally reduce on $S^1\times E_5$, where 
$S^1$ is parametrised by $\s$, to obtain a truncated $D=4$ theory of
gravity coupled to a vector and a scalar field.

We consider the type IIB ansatz
\begin{align}
ds^{2}=&ds^{2}_{4}+e^{2T}\left(d\s+A \right)^{2}+ds^{2}\left(SE_{5}\right)\nn
F_{5}=&4e^{T}\,\left(d\s
+A\right)\wedge \mathrm{Vol}_{4}+4\,\mathrm{Vol}\left(SE_{5}\right)\nn
H=&\,d\s\wedge W
\end{align}
with $W$ a harmonic (i.e. closed and co-closed) form on $SE_5$ satisfying
$\left|W\right|^{2}=4$, and trivial axion and dilaton. Here
the vector field $A$ and the scalar $T$ are defined on the four-dimensional
space corresponding to the line element $ds^2_4$.
The equations of motion of type IIB supergravity are satisfied 
provided that we satisfy the $D=4$ equations of motion
\begin{align}
R_{\mu\nu}=&-4g_{\mu\nu}+2A_{\mu}A_{\nu}+\nabla_{\mu}\nabla_{\nu}T+\partial_{\mu}T\partial_{\nu}T+\frac{1}{2}e^{2T}F_{\mu\lambda}F_{\nu}{}^{\lambda}\nn
 -\nabla^2T-\partial_{\mu}T\partial^{\mu}T=&-4+2e^{-2T}-\frac{1}{4}e^{2T}F_{\mu\nu}F^{\mu\nu}\nn
\nabla_{\nu}\left(e^{3T}F_{\quad\mu}^{\nu}\right)=&4e^{T}\,A_{\mu}\nn
A^{2}=&-e^{-2T}
\end{align}
These equations can be obtained from the $D=4$ action $S=\int d^4 x\sqrt{-g}{\cal L}$ with Lagrangian given by
\begin{align}
{\cal L}= & e^{T}\left[R +12-2e^{-2T}-\frac{1}{4}e^{2T}F_{\mu\nu}F^{\mu\nu}-2A^2\right]
\end{align}
provided that we impose the constraint $A^2=-e^{2T}$ by hand in the equations of motion.
Thus, the 
reduction does not comprise a ``consistent KK reduction'' in
the technical sense. In the next section we will see how such a reduction 
can be achieved for the case $E_5=T^{1,1}$. 

If we set $T=0$ the equations of motion can be derived
from the Lagrangian
\begin{align}
{\cal L}= & \left[R+10-\frac{1}{4}F_{\mu\nu}F^{\mu\nu}-2A^2\right]
\end{align}
describing a massive vector, with $mass^2=4$, coupled to gravity plus cosmological constant
as in \cite{Kachru:2008yh}\cite{Taylor:2008tg}, provided that we impose 
both $A^2=-1$ and $F_{\mu\nu}F^{\mu\nu}=-8$ by hand.

As a consistency check, 
one can directly verify that the Lif$_{4}$($z=2)$ solution 
\bea
ds^2_3&=&-r^{4}\,dt^{2}+r^{2}\,\left(dx_{1}^{2}+dx_{2}^{2}\right)+\frac{dr^{2}}{r^{2}}\nn
A&=&r^{2}\,dt
\eea
with $T=0$ solves the above equations of motion.

\subsection{Consistent KK reduction of type IIB on $S^1\times T^{1,1}$ to $D=4$}

We now consider the special case when the Einstein space is
$T^{1,1}$. For this case we can dimensionally reduce on
$T^{1,1}\times S^1$ to obtain an unconstrained $D=4$ theory of gravity coupled
to three scalar fields and a vector field. It is highly likely that
one can substantially extend this construction, probably consistent with supersymmetry,
(see \cite{Papadopoulos:2000gj}-\cite{Bena:2010pr}).
We expect these results to be important in future developments on these solutions.

We first introduce the notation for parametrising $T^{1,1}$
\begin{align}
ds_{i}^{2}=&\frac{1}{6}\,\left(d\theta_{i}^{2}+\sin^{2}\theta_{i}\,d\phi_{i}^{2}\right)\nn
J_{i}=&\frac{1}{6}\,\sin\theta_{i}\,d\theta_{i}\wedge d\phi_{i}\nn
D\psi=&d\psi-\cos\theta_{1}\,d\phi_{1}-\cos\theta_{2}\,d\phi_{2}
\end{align}
We then consider
 the type IIB reduction ansatz 
\begin{align}\label{lastansatz}
ds_{10}^{2}= & ds_{4}^{2}+e^{2T}\,\left(d\s+A\right)^{2}+e^{2V}\, \frac{1}{9}D\psi^{2}+e^{2U}\,\left(ds_{1}^{2}+ds_{2}^{2}\right)\nn
F_{5}= & 4e^{T-V-4U}\mathrm{Vol_{4}}\wedge\left(d\s+A\right)+\frac{4}{3}\, D\psi\wedge J_{1}\wedge J_{2}\nn
H= &\sqrt{2}\, \left(d\s+dk\right)\wedge\left(J_{1}-J_{2}\right)
\end{align}
with non-trivial ten dimensional dilaton $\phi$ and vanishing axion. 
Here $T$, $U$, $V$, $\phi$ and $k$ are scalars and $A$ is a vector field defined
on the four-dimensional space corresponding to the line element $ds^2_4$. 
%
We note that, if desired, one can remove the $dk$ term from $H$ by redefining $\sigma\to\sigma-k$, and this transformation
indicates that $k$ is a St\"uckelberg scalar for the vector field $A$, as we shall see below.
We also note that after $T$-duality on the $\s$ direction this then uplifts to the following
ansatz for $D=11$ supergravity
\begin{align}
ds_{11}^{2}=&e^{-2\left(\phi-T \right)/3}\left[ ds_{4}^{2}+e^{-2T}\,\left(d\s+
{\mathcal{A}}\right)^{2}+\frac{1}{9}e^{2V}\, D\psi^{2}+e^{2U}\,\left(ds_{1}^{2}+ds_{2}^{2}\right) \right]+e^{4\left(\phi-T \right)/3}\,d\chi^2\nonumber\\
F_{4}=&4e^{T-V-4U}\mathrm{Vol_{4}}+d\left[(A-dk)\wedge \left(d\s+\mathcal{A} \right)\right]\wedge d\chi\nn
d{\mathcal{A}}=&\sqrt{2}\,\left(J_{1}-J_{2}\right)
\end{align}

After substituting the type IIB ansatz \eqref{lastansatz}
into the type IIB equations of motion, we obtain $D=4$ equations
of motion (for the $D=11$ ansatz we obtain an equivalent set of equations).
Specifically, from the type IIB dilaton and three-form equation of motion we obtain
\begin{align}
&e^{-T-V-4U}\nabla_{\mu}\left(e^{T+V+4U}\,\nabla^{\mu}\phi\right)=-2e^{-\phi-4U}\,\left[e^{-2T}+\left(dk-A\right)^{2}\right]\nn
&d\left(e^{-\phi+T+V}\ast_{4}\left(A-dk\right)\right)=0
\end{align}

From the type IIB Einstein equations we obtain
\begin{align}
&6e^{-2U}-2e^{2V-4U}-\nabla^2 U-\partial_{\mu}T\partial^{\mu}U-4\partial_{\mu}U\partial^{\mu}U-\partial_{\mu}U\partial^{\mu}V=4e^{-2V-8U}\nonumber\\
&\qquad\qquad\qquad\qquad\qquad\qquad\qquad\qquad\qquad
+\frac{1}{2}e^{-\phi-4U}\left[e^{-2T}+\left(dk-A\right)^{2} \right]\nn
&4e^{2V-4U}-\nabla^2 V-\partial_{\mu}T\partial^{\mu}V-4\partial_{\mu}U\partial^{\mu}V-\partial_{\mu}V\partial^{\mu}V=4e^{-2V-8U}\nonumber\\
&\qquad\qquad\qquad\qquad\qquad\qquad\qquad\qquad\qquad
- \frac{1}{2} e^{-\phi-4U}\left[e^{-2T}+\left(dk-A\right)^{2}\right]\nn
& -\frac{1}{2}e^{-2T-V-4U}\nabla_{\mu}\left(e^{3T+V+4U}F^{\mu}{}_{\nu}\right)=2e^{-\phi-4U-T}\,\left(\partial_{\nu}k-A_{\nu} \right)\nn
& -\nabla^2 T-\partial_{\mu}T\partial^{\mu}T-4\partial_{\mu}T\partial^{\mu}U-\partial_{\mu}T\partial^{\mu}V+\frac{1}{4}e^{2T}F_{\mu\nu}F^{\mu\nu}=-4e^{-2V-8U}\nonumber\\
&\qquad\qquad\qquad\qquad\qquad\qquad\qquad\qquad\qquad
+\frac{1}{2}e^{-\phi-4U}\left[3e^{-2T}- \left(dk-A\right)^{2} \right]\nn
& R_{\mu\nu}-\nabla_{\mu}\nabla_{\nu}T-\partial_{\mu}T\partial_{\nu}T-\frac{1}{2}e^{2T}F_{\mu\lambda}F_{\nu}{}^{\lambda}-4\left(\nabla_{\mu}\nabla_{\nu}U+\partial_{\mu}U\partial_{\nu}U\right)\nonumber\\
& -\left(\nabla_{\mu}\nabla_{\nu}V+\partial_{\mu}V\partial_{\nu}V\right)=\frac{1}{2}\partial_{\mu}\phi\partial_{\nu}\phi-4e^{-2V-8U}g_{\mu\nu}\nonumber\\
&+\frac{1}{4}e^{-\phi}\left[8e^{-4U}\left(\partial_{\mu}k-A_{\mu} \right)\left(\partial_{\nu}k-A_{\nu} \right)-2g_{\mu\nu}e^{-4U}\left[e^{-2T}+\left(dk-A \right)^{2} \right] \right]
\end{align}

These equations can all be obtained from the 
$D=4$ action $S=\int d^4x{\sqrt{-g}}{\cal L}$ with Lagrangian given by 
\bea
{\cal L}&=&e^{T+4U+V}\Bigg[R
+12\,\partial U^{2}+8\partial U\partial V+8\partial T\partial U+2\partial T\partial V-\frac{1}{4}e^{2T}F_{\mu\nu}F^{\mu\nu}-\frac{1}{2}\partial\phi^{2}\nn
&+&24e^{-2U}-4e^{2V-4U}-8e^{-2V-8U}-2e^{-\phi-4U-2T}-2e^{-\phi-4U}(dk-A)^{2}\Bigg]
\eea
We can also rewrite this in the Einstein-frame by defining $g=e^{-T-4U-V}g_E$ and we obtain 
$S=\int d^4x{\sqrt{-g_E}}{\cal L}_E$
with 
\begin{align}
{\cal L}_E=&R_{E}+e^{-T-4U-V}\left(24e^{-2U}-4e^{2V-4U}-8e^{-2V-8U}-2e^{-\phi-4U-2T}\right)\nonumber\\&-2e^{-\phi-4U}\left(dk-A\right)^{2} -12\,\partial U^{2}-\frac{3}{2}\partial V^2-4\partial U\partial V-\frac{3}{2}\partial T^{2}-4\partial T\partial U\nonumber\\&-\partial T\partial V-\frac{1}{4}e^{3T+4U+V}F_{\mu\nu}F^{\mu\nu}-\frac{1}{2}\partial\phi^{2}
\end{align}

One can easily check that the equations of motion reduce to those consider in 
the last subsection after setting $\phi=U=V=k=0$.
Note that a key reason why this more general consistent KK reduction exists for
$T^{1,1}$ is that the form of
$W$ has the property that the non-zero components
of $W^2_{ij}$ are proportional to $\delta_{ij}$ on $T^{1,1}$. This will not be the
case\footnote{Although we note that it is also true for the non-supersymmetric 
case of $S^2\times S^3$ with $W$ proportional to the volume form
on $S^2$ and similarly for $T^{p,q}$.} for general $E_5$ and $W$.

\subsection{Reduction of $D=11$ on $S^1\times E_7$ to $D=3$}\label{a3}
We show how the special class of $D=11$ solutions of section \ref{d=11}, 
for arbitrary Einstein space, with $V$ independent of $\s$ and $f=1$
can be obtained from a $D=3$ of gravity after dimensional reduction
on $S^1\times E_7$.

Consider the following ansatz for $D=11$ supergravity
\bea
ds^2&=& ds^2_3+ \frac{e^{2T}}{4}(d\s+A)^2+ ds^2(E_7)\nn
G_4&=&3e^{T} d\s\wedge Vol_3+ \frac{1}{2}d\s\wedge V
\eea
Here $V$ is a harmonic (closed and co-closed) form on $E_7$ satisfying $|V|^2_E=8$
and $ds^2_3, A$ and $T$ are 
a three-dimensional metric, vector potential and scalar field, respectively.
We find that the $D=11$ equations of motion are satisfied provided that the following
$D=3$ equations are satisfied
\bea
R_{\mu\nu}&=&\nabla_{\mu}\nabla_{\nu}T+\nabla_{\mu}T\nabla_{\nu}T-12g_{\mu\nu}+\frac{1}{8}e^{2T}F_{\mu\rho}F_\nu{}^\rho+A_\mu A_\nu\nn
-\nabla^{2}T-\partial_{\nu}T\partial^{\nu}T&=&-12+4e^{-2T}-\frac{1}{16}e^{2T}F_{\mu\nu}F^{\mu\nu}\nn
\nabla_{\nu}\left(e^{3T}F_{\quad\mu}^{\nu}\right)&=&8e^{T}\,A_{\mu}\nn
A_\mu A^\mu&=&-4e^{-2T}
\eea
These equations can all be obtained from the 
$D=3$ action $S=\int d^3x{\sqrt{-g}}{\cal L}$ with Lagrangian given by 
\be
{\cal L}=e^{T}\left[R+24-4e^{-2T}-\frac{1}{16}e^{2T}F_{\mu\nu}F^{\mu\nu}-A^2\right]
\ee
provided that we impose the constraint $A_\mu A^\mu=-4e^{-2T}$ by hand in the equations of motion. 
If we set $T=0$ we have a theory of gravity with cosmological constant and a massive vector with $mass^2=8$,
but we have to now impose $A^2=-4$ and $F_{\mu\nu}F^{\mu\nu}=-128$ by hand.
One can directly check that  the Lif$_3$($z=2)$ solution in $D=3$
\bea
ds^2_3&=&\frac{1}{4}\left[-\rho^4 dt^2+\rho^2 dx^2+\frac{d\rho^2}{\rho^2}\right]\nn
A&=&\rho^2 dt
\eea
with $T=0$ solves these equations. 
We expect that more elaborate and consistent KK reductions can be made
for special choices of $E_7$ such as $E_7=S^3\times E_4$.

\section{Checking supersymmetry for Lif$_4$($z=2$) solutions in $D=11$}\label{susystructure}
We consider the $D=11$ ansatz
\begin{align}
ds_{11}^{2}=&-\Delta^{2}\,dt^{2}+\Delta^{-1}\,\left[H_{1}^{-1}\,\left(dx_{1}^{2}+dx_{2}^{2}\right)+H_{2}^{-1}\,\left(D\chi_{1}{}^{2}+D\chi_{2}{}^{2} \right)+ds^{2}\left(CY_{3}\right) \right]\nonumber\\
G_{4}=& dt\wedge d[J_{SU(5)}]\nonumber\\
\end{align}
where
\begin{equation}
\Delta\equiv  H_{1}^{-1/3}H_{2}^{-1/3},\qquad dD\chi_{i}=-W_{i}
\end{equation}
and the functions $H_{1}$, $H_{2}$ and the two-forms $W_{i}$ are defined on the Calabi-Yau three-fold
$CY_{3}$. 
The two-form $J_{SU(5)}$ is defined to be
\be
J_{SU(5)}=H_{1}^{-1}\,dx_{1}\wedge dx_{2}+H_{2}^{-1}\,D\chi_{1}\wedge D\chi_{2}+J_{CY}
\ee
where $J_{CY}$ is the K\"ahler-form on $CY_3$.

We demand that this is a supersymmetric solution of $D=11$ supergravity
with a time-like Killing spinor, by demanding that it has an $SU(5)$ structure
satisfying the conditions given in \cite{gp}. The $SU(5)$ structure is 
given by\footnote{Note that in \cite{gp} $J_{SU(5)}$, $\Omega_{SU(5)}$ were
denoted by $\Omega$, $\chi$, respectively.} 
$J_{SU(5)}$ and the $(5,0)$ form $\Omega_{SU(5)}$ defined by
\begin{align}
\Omega_{SU(5)}=&H_{1}^{-1/2}H_{2}^{-1/2}\,\left(dx_{1}+i\,dx_{2}\right)\wedge\left(D\chi_{1}+i\,D\chi_{2} \right)\wedge \Omega_{CY}
\end{align}
where $\Omega_{CY}$ is the holomorphic $(3,0)$ form on $CY_3$. 
The relevant conditions that need\footnote{In the language of \cite{gp} 
note that these imply, in particular, 
that $W_4=3d\log\Delta$ and $W_5=-12d\log\Delta$.}
to be imposed are \cite{gp}
\begin{align}
d\left(\Delta^{-3}\,[J_{SU(5)}]^4\right)=&0\nn
d\left(\Delta^{-3/2}\,\Omega_{SU(5)} \right)=&0
\end{align}
The first equation implies that ${W}=W_{1}+i W_{2}$ is primitive and the second one implies
that its type $\left(0,2\right)$ component is missing. In order that we solve the $D=11$ equations of motion
we need to also impose the equation of motion for the four-form which gives
\begin{align}
\nabla^2_{CY}H_{1}=&0\nn
\nabla^2_{CY}H_{2}=&-\left|{W}\right|^{2}
\end{align}

After the above remarks we observe see that our solution \eqref{eq:11D_susy} does indeed
lie within this class with
\begin{align}
ds^{2}(CY_3)=&dr^{2}+r^{2}\,ds^{2}\left(SE_{5}\right)\nn
H_{1}=&r^{-4}, \qquad H_{2}=fr^{-2}
\end{align}
with ${W}$ defined on $SE_5$ and $(1,1)$ and primitive on $CY_3$,
and $f$ satisfying equation \eqref{theeqagain}.

\end{document}